\documentclass[aps,prl,floatfix,twocolumn,groupedaddress,showpacs]{revtex4}

\usepackage{hyperref}
\usepackage{amsmath}
\usepackage{epsfig, graphicx}

\begin{document}
\newcommand{\simgt}{\,\hbox{\lower0.6ex\hbox{$\sim$}\llap{\raise0.6ex\hbox{$>$}}}\,}
\newcommand{\simlt}{\,\hbox{\lower0.6ex\hbox{$\sim$}\llap{\raise0.6ex\hbox{$<$}}}\,}

\title{Temperature Dependence of Electric Field Noise Above Gold Surfaces}

\author{Jaroslaw Labaziewicz}
\email[]{labaziew@mit.edu}
\author{Yufei Ge}
\author{David R. Leibrandt}
\author{Shannon X. Wang}
\author{Ruth Shewmon}
\author{Isaac L. Chuang}
\affiliation{Massachusetts Institute of Technology, Center for Ultracold Atoms, Department of Physics, 77 Massachusetts Avenue, Cambridge, MA, 02139, USA}

\date{\today}

\begin{abstract}
Electric field noise from fluctuating patch potentials is a significant problem for a broad range of precision experiments, including trapped ion quantum computation and single spin detection.
Recent results demonstrated strong suppression of this noise by cryogenic cooling, suggesting an underlying thermal process.
We present measurements characterizing the temperature and frequency dependence of the noise from $7$ to $100~\text{K}$, using a single $\text{Sr}^+$ ion trapped $75~\mu\text{m}$ above the surface of a gold plated surface electrode ion trap.
The noise amplitude is observed to have an approximate $1/f$ spectrum around $1$ MHz, and grows rapidly with temperature as $T^\beta$ for $\beta$ from $2$ to $4$.
The data are consistent with microfabricated cantilever measurements of non-contact friction but do not extrapolate to the DC measurements with neutral atoms or contact potential probes.
\end{abstract}

%\keywords{}

\maketitle

The surface of a metal is ideally an electrical equipotential, but in reality it exhibits significant potential variations, up to hundreds of millivolts over micrometer distances.
These ``patch potential'' variations generate local electric fields, with a static component thought to originate from differences in the work function between crystal facets, further modified by adsorbates.
Because of the impact these fields can have on precision measurements, static patch fields have been extensively characterized in studies of neutral atoms\cite{Grabka:87, Hinds:93, Cornell:07}, gravitational forces\cite{Fairbank:77, Zhou:06}, electron emission\cite{Nichols:49} and contact potentials\cite{Brown:91, Opat:92, Bsiesy:06} for a wide variety of materials and morphologies.

Patch potentials also fluctuate in time, a process about which surprisingly little is known, but one with broad practical implications in trapped ion quantum computation\cite{James:98b, Turchette:00, Deslauriers:04, Seidelin:06}, nanomechanics\cite{Gotsmann:99, Rugar:01, Marohn:06}, single spin detection\cite{Rugar:03} and measurements of weak forces\cite{Trenkel:03, Jamell:05}.
The origin of the fluctuations is unknown, but one startling observation has been that cooling of the metal surfaces suppresses the noise by many orders of magnitude, in both ion traps\cite{Deslauriers:06, Labaziewicz:08} and microfabricated cantilevers\cite{Rugar:01}.
One might expect that this suppression originates from activation barriers for processes changing the surface potential, such as diffusion of surface adsorbates.
However, so far there is no systematic data on the dependence of the fluctuation amplitude on temperature, which could support or refute such a hypothesis, illuminating the underlying physical mechanism.

Here, we present a controlled study of the temperature dependence of the amplitude of electric field noise between $0.6~\text{MHz}$ and $1.5~\text{MHz}$, and from $7$ to $100~\text{K}$, measured by a single trapped ion located $75~\mu\text{m}$ above a gold surface.  We begin, below, with a description of the surface electrode trap used in the experiment, and summarize the strategies employed to control errors.
Eight datasets are presented, showing a noise amplitude which grows with temperature as $T^{\beta}$ where $2\simlt \beta \simlt 4$.
The spectrum is found to scale with frequency as $f^{-\alpha}$ with $\alpha\sim 1$ at $T = 100\text{K}$, but $\alpha\simlt 1$ for $T<100$ K.
We show in the analysis that this temperature dependence is inconsistent with expectations from the previously proposed theoretical models based on thermal voltage fluctuations\cite{Lamoreaux:97, Henkel:99a}, and propose as an alternative a model based on a continuous spectrum of thermally activated random processes.
We conclude by noting consistency between our observed noise amplitudes and temperature dependence with measurements using cantilevers, but find that the field amplitudes observed at $\sim 1~\text{MHz}$ do not extrapolate well to the DC amplitudes observed in other systems.

%\section{Experiment Design}
The trap design is identical to the smallest of the traps described in Ref.~\cite{Labaziewicz:08}, with the trap center $75~\mu\text{m}$ above the surface.
The fabrication process is similar to that of Ref.~\cite{Seidelin:06}, and is described only briefly here.
Initially, a $10~\text{nm}$ layer of Ti, followed by a $100~\text{nm}$ layer of Ag, is evaporated onto a crystal quartz substrate, chosen for its high thermal conductivity at cryogenic temperatures.
Trap electrodes are patterned with AZ 4620 photoresist, and the
exposed areas electroplated using Transene \mbox{TSG-250} gold plating solution.
A $1~\text{k}\Omega$ heating element and two $\text{RuO}_2$ temperature sensors, on opposite sides of the trap, are soldered to the surface of the trap.
The finished chip is glued in a ceramic pin grid array carrier with a low vapor pressure epoxy (TorrSeal), cleaned in laboratory solvents, dried at $100~^\circ\text{C}$, exposed to a UV/ozone lamp to remove organic residue and transferred to the vacuum chamber within hours of cleaning.
Given the strong dependence of noise on fabrication process observed in previous experiments\cite{Labaziewicz:08}, we fabricated $4$ traps.
The room temperature heating rates were measured in one of these traps (trap IV) to be $4200~\pm~300$ quanta/s at $1~\text{MHz}$, in good agreement with published data\cite{Seidelin:06, Epstein:07}.

The trap is cooled by contact to the $4~\text{K}$ plate of a helium bath cryostat.
The poor thermal conductivity of the ceramic carrier allows for stabilization of the trap temperature at $7$ to $100~\text{K}$, limited by the maximum power dissipated by the heating element ($500~\text{mW}$).
The RF and DC sources, leads and filters remain at a constant temperature throughout our measurement, contributing at most a temperature independent noise offset.

To measure the field noise, a single $\text{Sr}^+$ ion, produced by photoionization of a thermal vapor, is loaded into the trap.
The ion is Doppler cooled to $<1~\text{mK}$, then the lowest frequency vibrational mode is sideband cooled to the motional ground state, with better than $99\%$ probability.
The number of phonons, $n$, in this mode can be derived from the transition probabilities on the blue ($P_{bsb}$) and red motional sidebands ($P_{rsb}$) of the $S_{1/2} \leftrightarrow D_{5/2}$ transition using $n = \frac{P_{rsb}}{P_{bsb} - P_{rsb}}$\cite{Turchette:00}.
The heating rate $\dot{n}$, obtained by introducing a variable delay between sideband cooling and measurement of the transition probabilities, is converted to electric field noise using
\[
S_{E}(f) = \frac{4 m \hbar \left(2 \pi f\right)}{q^2} \dot{n} \approx 15~\dot{n}~\frac{f}{1~\text{MHz}}\times10^{-15}~\text{V}^2/\text{m}^2/\text{Hz}
\]
where $m$ is the ion mass, $f$ the trap frequency and $q$ the charge of the ion.
%The scan is repeated 10 times, and the statistic of these measurements is used to derive experimental error bars.
%Laser drifts are eliminated by locking to the transition using Ramsey spectroscopy with a $90^\circ$ phase shift between the two $\pi/2$ pulses\cite{Sinclair:04}.

In order to minimize other sources of heating, all DC sources are low-pass filtered at $4~\text{kHz}$, and the RF source is high-pass filtered at $10~\text{MHz}$.
Ion micromotion is reduced using the photon-correlation method to $<10~\text{nm}$ in the plane of the trap and $<100~\text{nm}$ perpendicular to that plane\cite{Berkeland:98}.
The observed heating rates do not depend strongly on DC and RF voltages near the optimal operating point.
Electric field noise data was taken in a non-sequential order of temperatures to guard against remaining temperature independent sources.
Finally, values measured at $6~\text{K}$ increased by $5\pm4~\times10^{-14}~\text{V}^2/\text{m}^2/\text{Hz}$ after loading $\approx 1000$ ions.
Typically, fewer than 50 ions are loaded over a trap lifetime, and therefore $\text{Sr}$ contamination is not expected to affect our noise measurements.
For more details about the measurements, see Ref.~\cite{Labaziewicz:thesis}.

\begin{figure}\includegraphics[width=3.4in]{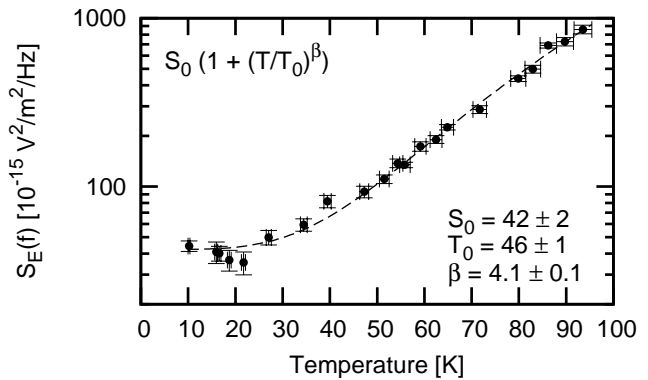}
\caption{\label{Fig:TempDep} Temperature dependence of the measured field noise in trap II (dots) and the fit to $S_E(T) = 42\left(1+\left(T/46~\text{K}\right)^{4.1}\right) \times 10^{-15}~\text{V}^2/\text{m}^2/\text{Hz}$ (dashed line).}
\end{figure}

\begin{figure}\includegraphics[width=3.4in]{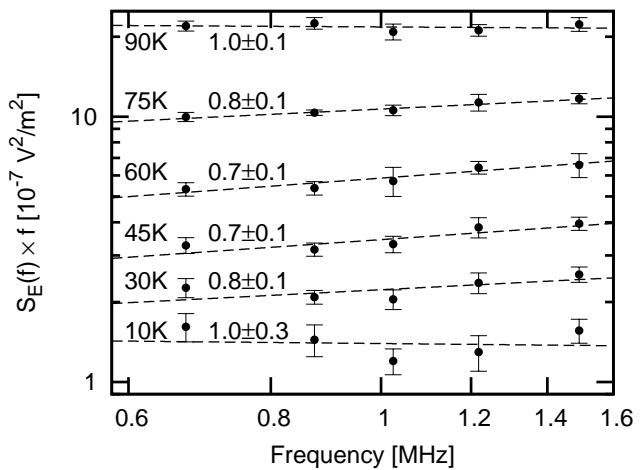}
\caption{\label{Fig:FreqDep} Noise spectrum at $6$ different temperatures measured in trap III a.
Plotted are the values of $S_E(f) \times f$ and fit lines to $S_E(f) \times f = f^{1-\alpha}$.
The fit exponent $\alpha$ and the measurement temperature are indicated above fit lines.}
\end{figure}

%\section{Temperature and Frequency Dependence of Noise}
The temperature dependence was characterized in four traps, with one trap measured five separate times, resulting in $8$ datasets.
Except where noted, data was taken at $0.84$ to $0.88~\text{MHz}$ and scaled to $1~\text{MHz}$ assuming $1/f$ scaling.
The observed field noise remains rather flat below $40~\text{K}$, but increases rapidly with temperature above that point, where it can be modelled accurately by a polynomial $T^\beta$ (Fig.~\ref{Fig:TempDep}).
Table~\ref{Tbl:Data} summarizes the parameters of fits of the datasets to
the experimentally motivated form $S_E(T) = S_0 \left(1+(T/T_0)^\beta\right)$.
The frequency spectrum of the noise was measured in one trap (trap III a), and fitted to $f^{-\alpha}$.
At $100~\text{K}$, the exponent $\alpha$ is very close to $1$, consistent with published values\cite{Deslauriers:06, Epstein:07}.
At lower temperatures, however, $\alpha \approx 0.7$ (Fig.~\ref{Fig:FreqDep}).

\begin{table*}\centering\begin{tabular}{| l | c | c | c | c |}
\hline
& \multicolumn{3}{ c |}{Fit Parameters} & \\
Trap & $S_0~[10^{-15}~\text{V}^2/\text{m}^2\text{Hz}]$ & ~~~~ $T_0~\text{[K]}$ ~~~~ & ~~~~ $\beta~[1]$ ~~~~ & Notes \\ \hline
~I      & $65 \pm 3$ & $73 \pm 3$ & $3.0 \pm 0.2$ & $6^\text{th}$ cooldown\\ \hline
~II     & $42 \pm 2$ & $46 \pm 1$ & $4.1 \pm 0.1$ & Initial cooldown \\ \hline
~III a) & $167 \pm 7$ & $46 \pm 1$ & $3.6 \pm 0.2$ & Initial cooldown \\ \hline
~~~~ b) & $120 \pm 10$ & $45 \pm 3$ & $3.5 \pm 0.2$ & Temperature cycle to 130K while in vacuum \\ \hline
~~~~ c) & $54 \pm 3$ & $44 \pm 2$ & $3.2 \pm 0.1$ & Temperature cycle to 340K while in vacuum \\ \hline
~~~~ d) & $60 \pm 4$ & $49 \pm 4$ & $2.1 \pm 0.1$ & Recleaning in lab solvents in air \\ \hline
~~~~ e) & $18 \pm 3$ & $17 \pm 3$ & $1.8 \pm 0.1$ & Recleaning in lab solvents in air \\ \hline
~IV     & $3300 \pm 40$ & $73 \pm 1$ & $3.2 \pm 0.1$ &~Following the room temperature measurements~\\ \hline
\end{tabular}
\caption{\label{Tbl:Data} Summary of the fit parameters to $S_E(T) = S_0 \left(1+(T/T_0)^\beta\right)$ obtained from the measurements. 
Trap III was measured five times, with in-between processing steps indicated in the notes column.
Trap IV was measured in a room temperature system prior to measurements at cryogenic temperatures.}
\end{table*}

The exponent $\beta$ varies between the traps and separate measurements in the same trap.
In trap III, the observed noise amplitude was stable for hours (III a), before abruptly dropping after temperature cycling to $130~\text{K}$ (III b).
Temperature cycling to $340~\text{K}$, without breaking vacuum, resulted in a further decrease (III c).
Exposure to air and laboratory solvents did not return the observed fluctuations to initial values (III d,e), indicating that temperature cycling results in an irreversible change.
The lowest obtained value (III e) is more than $2$ orders of magnitude below best reported values for similarly sized traps operated at room temperature\cite{Epstein:07}.
Trap IV, used to measure the noise at room temperature, exhibited significantly higher heating rates at cryogenic temperatures, possibly related to temperature cycling between $77~\text{K}$ and $200~^\circ\text{C}$ or cleaning in hot acetone, performed when removing the trap from the room temperature system.
Despite strong variation in the amplitudes $S_0$, the turn-on temperatures $T_0$ are consistent through the dataset.
The finite zero-temperature intersect indicates either zero temperature fluctuations, or a thermally activated process with activation energy less than $7~\text{K}$\cite{Schlosshauer:08}.

%\section{Anomaly}

\begin{figure}
\includegraphics[width=3.4in]{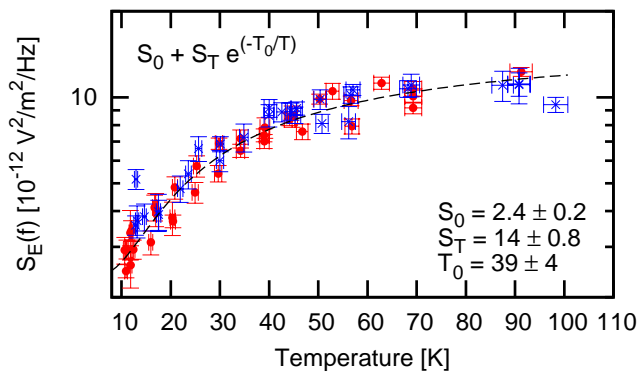}
\caption{\label{Fig:Anomaly} (color online) Anomalous behavior observed in trap II, after cleaning in an ultrasonic bath. Datapoints taken at $0.86~\text{MHz}$ (red circle) and $1.23~\text{MHz}$ (blue $\times$) were scaled to $1~\text{MHz}$ assuming $1/f$ scaling and fitted to an Arrhenius curve (dashed line).}
\end{figure}

In one instance, we found a dramatically distinct behavior.
After trap II was cleaned in an ultrasonic bath, the surface became visibly roughened.
The observed heating rates increased from the initial values by $2$ orders of magnitude.
The temperature dependence could not be fitted to the $T^\beta$ form, but followed an Arrhenius curve, $S_E(T) = S_0 + S_T e^{-T_0/T}$, with activation energy $T_0 \approx 40~\text{K}$ (Fig.~\ref{Fig:Anomaly}).
Good agreement of the fit and experimental data suggests that the
noise is dominated by a narrow range of activated
processes\cite{Connors:90}.

%\section{Discussion of Data}

The measured temperature dependences allow us to examine the viability of proposed theoretical models of the noise.
In particular, thermal voltage fluctuation (Johnson-like) models scale as $\rho\times T$, where $\rho$ is the resistivity of the metal electrodes\cite{Lamoreaux:97, Henkel:99a}.
Given the almost linear dependence of gold resistivity on temperature
above $20~\text{K}$, such a model would predict a $T^2$ dependence, at odds with the observed exponents up to $4.1$.
Similarly, the high exponent suggests that the mechanism responsible for heating in ion traps is fundamentally distinct from that responsible for charge noise in condensed matter systems, where observed noise scales as $T$ or $T^2$\cite{Clarke:04, Tsai:06, Mooij:06}.

The observed $T^{\beta}$ dependence can be explained by assuming a spectrum of thermally activated random processes (charge traps, adsorbate diffusion etc.), with density of activation energies $D(E) \propto E^{\beta - 1}$\cite{DuttaHorn:81}.
Such a model naturally accounts for the $f^{-\alpha}$ frequency
dependence with a temperature dependent exponent given by
\[
\alpha = 1 - \frac{1}{\ln(\omega \tau_0)} \left( \frac{\beta (T/T_0)^\beta}{1 + (T/T_0)^\beta} - 1 \right)
\]
where $\tau_0$ is the shortest timescale in the system, assumed to be
$10^{-12}~\text{s}$ (inverse phonon frequency).  This model predicts
that $\alpha$ should be slightly less than $1$ when
$T < T_1 = T_0 / (\beta - 1)^{1 / \beta} \approx 35~\text{K}$ (trap III a) and slightly more than $1$ when $T > T_1$.
While our data exhibits qualitatively similar behavior, 
experimentally $\alpha$ remains below $1$ up to $T \approx 90~\text{K}$.
A further refinement of the model would be necessary to resolve this issue.

%\section{Extrapolation and Conclusion}

It is instructive to compare the measured electric field noise with what is observed in other systems, assuming the surface distance and frequency scalings to be $d^{-4}$ and $f^{-1}$, as observed in ion traps\cite{Turchette:00, Deslauriers:06}.
Rewriting the room temperature data as $S_E(f) = f^{-1} d^{-4}\times10^{-21}\text{V}^2\text{m}^2$, the expected noise at $100~\text{nm}$ and $10~\text{kHz}$ is $S_E \approx 10^3~\text{V}^2/\text{m}^2/\text{Hz}$.
Refs.~\cite{Gotsmann:99, Marohn:06} use the damping rate $\Gamma$ of the mechanical motion of a microfabricated cantilever to measure the electric field noise above a gold surface as $S_E(f) = (4 k_B T \Gamma) / (C V)^2$.
Here $k_B$ is the Boltzmann constant and $C$ and $V$ are the capacitance and voltage between the surface and the cantilever tip.
Both papers use cantilevers tips with a radius of curvature $\rho \simlt 30 \textrm{nm}$ and report electric field noise within an order of magnitude of that extrapolated from this work.
Ref.~\cite{Rugar:01} makes a similar measurement using a cantilever tip with $\rho \approx 1 \mu\textrm{m}$ and reports noise field suppression by a factor of $20$ upon cooling to $77~\text{K}$ and an additional factor of $25$ after cooling to $4~\text{K}$, within a factor of $2$ of the temperature dependence observed in this work.
The amplitude of the field noise, however, is three orders of magnitude smaller than that extrapolated from this work; this is perhaps a consequence of field suppression in the regime $d \ll \rho$.
Reported distance scaling varies from $d^{-4}$ in the regime $d \approx \rho$\cite{Gotsmann:99} to $d^{-1}$ in the regime $d \ll \rho$\cite{Rugar:01}; Ref.~\cite{Marohn:06} reports a distance scaling in between.
Such probe size effects do not influence measurements with strongly confined ions, but have to be accounted for in a comparison of cantilever and ion trap experiments.
%Ref.~\cite{Gotsmann:99} is in the regime $\rho \approx d$ and reports distance scaling $d^{-4}$ while Ref.~\cite{Rugar:01} is in the regime $\rho \gg d$ and reports distance scaling $d^{-1}$; Ref.~\cite{Marohn:06} reports distance scaling in between.
%A careful comparison of the noise in cantilever and ion trap experiments should take this possible field suppression and resulting distance scaling into account.

The results presented can also be extrapolated to very low frequencies by integrating the $1/f$ noise spectrum, allowing for a comparison with the measurements of static patch fields.
The extrapolated room temperature fluctuation of electric field, $\sigma_E$, in time $\tau = 1~\text{s}$ is approximately $\sigma_E^2 = S_E(f)\times f \times \ln\left(\tau/\tau_0\right) \approx d^{-4}\times10^{-20}~\text{V}^2\text{m}^2$\cite{Bobbert:97}.
For a surface distance $d = 1~\mu\text{m}$, we get $\sigma_E \approx 10^2~\text{V}/\text{m}$.  For the same surface distance, Eq.~(4) of Ref.~\cite{Hinds:93} reports a static field variance with $d^{-4}$ dependence of $\sigma_E \approx 10^4~\text{V}/\text{m}$.
This mismatch becomes worse for $d > 1~\text{cm}$ where the extrapolated fields become insignificant compared to the measured values\cite{Grabka:87}.
Contact potential experiments provide a different physical characteristic of the static patch fields, the product of the patch size $A_{patch}$ and the surface potential variance $\sigma_V^2$, with a reported value of $\sigma_V^2 A_{patch} \approx 10^{-12}~\text{V}^2\text{m}^2$\cite{Brown:91}.
The same quantity can be estimated from the field noise measurements presented here using $S_E(f) d^4 \approx S_V(f) A_{patch}$\cite{Turchette:00}.
Integrating over the frequency spectrum, we get $\sigma_V^2 A_{patch} \approx \sigma_E^2 d^4 \approx 10^{-20}~\text{V}^2\text{m}^2$.
One possible explanation of the mismatch between the extrapolation of this work to low frequencies and the direct measurements of Refs.~\cite{Grabka:87, Hinds:93, Brown:91} is that only a fraction of the static field sources fluctuate.

In conclusion, we have measured the temperature and frequency dependence of electric field noise above a gold surface using a single $\text{Sr}^+$ ion.
Noise field measurements show significant variability, including a 10-fold reduction after repeated cycling between $4$ and $300~\text{K}$, and a dramatic change in behavior upon cleaning in an ultrasonic bath.
The temperature dependence we observe is inconsistent with the previously proposed thermal voltage fluctuation models, but may be explained using a spectrum of thermally activated random processes.
Further work is needed to obtain quantitative agreement between this model and our observed frequency dependence.
Experimental noise amplitudes and temperature scaling are similar to those measured using microfabricated cantilevers, suggesting a common origin.
We hope that future improvements in understanding the spread of measured values in ion trap experiments and distance scaling in cantilever experiments will strengthen the correspondence.
% between electric field noise observed in ion traps and cantilevers.

\begin{acknowledgments}
We thank K. R. Brown and V. Vuletic for helpful advice. 
This work is supported in part by the Japan Science and Technology
Agency, the NSF CUA, and the ARO.
\end{acknowledgments}

% \bibliography{refs}

\end{document}